\documentclass[12pt,a4paper]{article}
\usepackage[usenames,dvipsnames]{pstricks} 
\usepackage{afterpage}
\usepackage{epsfig}
\usepackage{bmpsize}
\usepackage{pst-grad} 
\usepackage{pst-plot} 
\usepackage[a4paper]{geometry}
\usepackage{float}
\usepackage[stable]{footmisc}
\usepackage[utf8x]{inputenc}
\usepackage{a4wide}
\usepackage{amsmath,amssymb,amstext,amsthm,amssymb}
\usepackage{amsfonts}
\usepackage{framed}
\usepackage{graphicx} 
\usepackage{subcaption}
\usepackage{setspace}
\usepackage{bbold}
\usepackage{bbm}
\usepackage{amstext}
\usepackage{array}
\usepackage{hyperref}
\usepackage{overpic}
\usepackage{bbm}
\usepackage[nosort]{cite}
\usepackage{lipsum}
\usepackage{color}
\usepackage{enumitem}
\usepackage{cancel}
\usepackage{verbatim}

\graphicspath{{figures/}}

\newcommand{\del}[0]{\partial}

\let\baraccent=\=
\renewcommand{\=}[1]{\stackrel{#1}{=}}

\newcommand{\Tr}[1]{\text{Tr}\left( #1 \right)}

\newcommand{\IR}{\text{IR}}

\begin{document}
\pagestyle{plain}

\makeatletter
\@addtoreset{equation}{section}
\makeatother
\renewcommand{\theequation}{\thesection.\arabic{equation}}
\pagestyle{empty}
\rightline{DESY 19-012}
\vspace{0.5cm}
\begin{center}
\Huge{{Gaugino condensation \\and small uplifts in KKLT}
\\[15mm]}
\normalsize{Federico Carta,$^1$ Jakob Moritz,$^{1}$ and Alexander Westphal$^1$ \\[10mm]}
\small{\slshape
$^1$ Deutches Electronen-Synchrotron, DESY, Notkestra$\beta$e 85,\\ 22607 Hamburg, Germany. \\[2mm]  } 
\normalsize{\bf Abstract} \\[8mm]
\end{center}
\begin{center}
\begin{minipage}[h]{15.0cm} 

In the first part of this note we argue that ten dimensional consistency requirements in the form of a certain tadpole cancellation condition can be satisfied by KKLT type vacua of type IIB string theory. We explain that a new term of non-local nature is generated dynamically once supersymmetry is broken and ensures cancellation of the tadpole. It can be interpreted as the stress caused by the restoring force that the stabilization mechanism exerts on the volume modulus. In the second part, we explain that it is surprisingly difficult to engineer sufficiently long warped throats to prevent decompactification which are also small enough in size to fit into the bulk Calabi-Yau (CY). We give arguments that achieving this with reasonable amount of control may not be possible in generic CY compactifications while CYs with very non-generic geometrical properties might evade our conclusion.

\end{minipage}
\end{center}
\newpage
\setcounter{page}{1}
\pagestyle{plain}
\renewcommand{\thefootnote}{\arabic{footnote}}
\setcounter{footnote}{0}


\tableofcontents


\section{Introduction}
It is believed that of all the seemingly consistent effective field theories (EFT) coupled to gravity, only a small subset can actually arise as a low energy limit of a consistent theory of quantum gravity (usually assumed to be string theory) \cite{Vafa:2005ui}. The subset realizable in string theory is referred to as the \textit{the landscape} and the one impossible to realize as \textit{the swampland}. A natural question to ask is therefore whether a given EFT coupled to Einstein gravity belongs to the landscape or to the swampland, and addressing such questions is the aim of the swampland program.

Several criteria that may be able to distinguish those EFTs coupled to gravity that can be realized in string theory from those that cannot have been formulated. Such criteria include for example the weak gravity conjecture \cite{ArkaniHamed:2006dz,delaFuente:2014aca,Rudelius:2015xta,Brown:2015iha,Bachlechner:2015qja,Hebecker:2015rya,Harlow:2015lma,Heidenreich:2016aqi,Crisford:2017gsb,Lee:2018urn,Reece:2018zvv,Montero:2018fns,Lee:2019tst}, the distance conjecture \cite{Ooguri:2006in,Klaewer:2016kiy,Blumenhagen:2017cxt,Grimm:2018ohb,Heidenreich:2018kpg,Lee:2018spm,Grimm:2018cpv,Buratti:2018xjt,Gonzalo:2018guu}, instability of non-supersymmetric anti de-Sitter space (AdS) \cite{Ooguri:2016pdq,Ibanez:2017kvh}, finiteness of the number of massless fields, and various others \cite{Ooguri:2006in}. For a recent review see \cite{Brennan:2017rbf}.

The most recently proposed criterium \cite{Obied:2018sgi, Agrawal:2018own, Garg:2018reu, Ooguri:2018wrx} states that all the EFTs coupled to gravity with de Sitter (dS) vacua reside in the Swampland.\footnote{For a discussion regarding the form and viability of the conjecture we refer the reader to references \cite{Denef:2018etk,Roupec:2018mbn,Conlon:2018eyr,Akrami:2018ylq,Murayama:2018lie,Choi:2018rze,Hebecker:2018vxz}} This criterium is referred to as the \textit{no-dS conjecture}.

The \textit{no-dS conjecture} is in stark contrast to what an effective field theorist would conclude from the observational fact that our universe is undergoing accelerated expansion. Arguably, from her point of view a tiny but positive cosmological constant would be the simplest and most natural fit to the data. In particular, she would conclude that in the far future the geometry of our universe will be well described by a patch of de Sitter (dS) space. 

Some of the evidence for the conjecture comes from the fact that in various classical corners of string theory there exist no-go theorems against the existence of de Sitter solutions \cite{Maldacena:2000mw,deAlwis:2003sn,Hertzberg:2007wc,Flauger:2008ad,Danielsson:2009ff,Shiu:2011zt,Danielsson:2011au,Kutasov:2015eba,Andriot:2016xvq,Andriot:2017jhf}. Due to the existence of these theorems (and also due to the Dine Seiberg problem \cite{Dine:1985he}) it seems natural to expect that if any dS solutions exist at all, they will require a competition between classical and quantum effects that cannot be mapped into a purely classical effect by any duality transformation.\footnote{Whether this intuition is correct remains an open question as there are many proposals for purely classical meta-stable dS solutions, e.g. supercritical strings~\cite{Silverstein:2001xn,Maloney:2002rr,Dodelson:2013iba}, type IIB string theory compactified on orientifolded products of Riemann surfaces~\cite{Saltman:2004jh}, and more recent work in the context of F-theory~\cite{Polchinski:2009ch,Dong:2010pm,Dong:2011uf}. On the type IIA side there were studies of dS on (generalizations of) twisted tori~\cite{Silverstein:2007ac,Flauger:2008ad,Danielsson:2011au,Roupec:2018mbn}. These proposals should be further scrutinized in the future as they form possible counter examples against the no-dS conjecture.} One of the most convincing proposals of such a kind has been made in \cite{Kachru:2003aw} by Kachru, Kallosh, Linde and Trivedi (KKLT) (see section \ref{sec:KKLTsummary} for a summary).\footnote{For further dS proposals involving balancing classical and quantum effects, see e.g.~\cite{Burgess:2003ic,Balasubramanian:2004uy,Saltman:2004jh,Balasubramanian:2005zx,Conlon:2005ki,Westphal:2006tn,Cicoli:2012fh,Louis:2012nb,Cicoli:2013cha,Rummel:2014raa,Cicoli:2015ylx,Braun:2015pza,Retolaza:2015nvh,Gallego:2017dvd,Roupec:2018mbn}, and the recent review~\cite{Cicoli:2018kdo}. For a new perspective on dark energy from F-theory, see \cite{Heckman:2018mxl,Heckman:2019dsj}.} In this note we will focus on this model. We do so because 1) it is one of the most studied models \cite{Baumann:2006th,Baumann:2010sx,Dymarsky:2010mf}, 2) it is consistent with the general expectation that de Sitter vacua are non-generic and at best meta-stable \cite{Dine:1985he,Goheer:2002vf,Kachru:2003aw,Freivogel:2008wm}, and 3) there is evidence that the tuning requirements that make it non-generic can actually be met \cite{Denef:2008wq}.

The construction has recently been criticized\footnote{For a qualitatively different criticism we refer the reader to ref. \cite{Sethi:2017phn}.} from a ten-dimensional point of view \cite{Moritz:2017xto,Gautason:2018gln} that we will summarize in section \ref{sec:tadpole_cancellation}. The basic point is that ten-dimensional consistency requirements in the form of tadpole cancellation conditions seem to be in conflict with the existence of four-dimensional KKLT type de Sitter vacua. This criticism is based on the conjecture that the $10d$ lift of these vacua is accurately described by inserting by hand a non-vanishing expectation value for a certain gaugino bilinear $\langle \lambda\lambda \rangle \neq 0$. This conjecture is supported for instance by the fact that the non-perturbative superpotential for $D3$-brane position moduli can be accurately computed from ten dimensional supergravity via the insertion of the gaugino bilinear as a classical source term \cite{Baumann:2010sx}, and the ability to find supersymmetric backreacted solutions \cite{Dymarsky:2010mf}. We will call this the \textit{$10d$ gaugino condensation conjecture}. If this conjecture were true, roughly speaking, it would allow to constrain de Sitter vacua of the KKLT type much via the same tools that are used to exclude purely classical solutions in \cite{Maldacena:2000mw}.

The content of this note can be split into two more or less independent parts. In the first part we explain that while the $10d$ gaugino condensation conjecture can be argued to be valid for the supersymmetric KKLT AdS vacua, it will fail to hold once supersymmetry is broken: there are additional contributions to the $10d$ tadpole that can be shown to arise from demanding \textit{only} the consistency of the supersymmetric KKLT construction and that become relevant only once SUSY is broken. We find it unlikely that these contributions can be captured by a local $10d$ action in the above sense. Moreover, under the assumption that arbitrarily strongly warped regions exist in the flux compactification, these new contributions can be shown to precisely cancel the tadpole once SUSY is broken by a warped uplift. In total, we see no further reason to expect the failure of KKLT uplifts from considerations of $10d$ tadpole cancellation.

In the second part, we contrast this by arguing that a successful uplift to a dS vacuum via warped uplifts, and more so to a SUSY breaking AdS vacuum, is highly constrained by the geometrical consistency requirement that the warped throat used for the uplift must fit into the bulk CY.\footnote{Note that the same requirement has been used to constrain inflationary models in \cite{McAllister:2008hb} and in \cite{Freivogel:2008wm} to argue that the flux superpotential must be tuned extremely small for the KKLT construction to be consistent.} We will show that the simplest examples with a single K\"ahler modulus can hardly satisfy this basic requirement. For the case of many K\"ahler moduli we speculate that this problem becomes even more severe unless the compactification manifold satisfies additional geometrical properties that we believe are highly non-generic. It would be very interesting to investigate whether such geometries can be realized in a controlled manner.

This note is organized as follows. In section \ref{sec:KKLTsummary} we briefly review the KKLT proposal. In section \ref{sec:Gauginos} we give a simple minded motivation for why it may be consistent to treat gaugino condensates as classical sources for \textit{some} physical questions while this approach must fail for others. In section \ref{sec:tadpole_cancellation} we review the $10d$ tadpole cancellation puzzle and propose how to resolve it. In section \ref{sec:warped_uplifts} we show that realizing a sufficiently small warped uplift is impossible in the simplest examples and at least very much non-generic in the general case. We conclude in section \ref{sec:conclusions}.

\section{KKLT: A summary}\label{sec:KKLTsummary}


The KKLT proposal operates in the flux landscape of $O3/O7$ orientifolds of type IIB string theory in its $10d$ SUGRA approximation as pioneered by Giddings, Kachru and Polchinski (GKP) \cite{Giddings:2001yu}. The complex structure moduli $z^i$ and the axio-dilaton $\tau$ are stabilized via fluxes at a high scale (but below the KK-scale) via the Gukov-Vafa-Witten (GVW) superpotential \cite{Gukov:1999ya}
\begin{equation}\label{eq:GVWsuperpotential}
W_{GVW}=\int_{CY} G_3\wedge \Omega\, ,
\end{equation}
where $\Omega$ is the holomorphic three form of the CY, and $G_3=F_3-\tau H_3$ is the complex three-form. The GVW superpotential is independent of the K\"ahler moduli. The number of K\"ahler moduli is given by $h^{1,1}_+$ which is the dimension of the orientifold-even Dolbeault cohomology group $H^{1,1}_+(CY,\mathbb{R})$. For $h^{1,1}_+=1$, and after integrating out the complex structure moduli, the effective SUGRA of the single K\"ahler modulus $T$ reads
\begin{equation}\label{eq:KKLT_Kahler_superpotential_classical}
K=-3\log(T+\bar{T})\, ,\quad W=W_0=const.
\end{equation}
Here, $W_0$ is the value of the GVW superpotential evaluated at the frozen values of the $\{z^i,\tau\}$. This SUGRA model is valid to lowest order in the $\alpha'$ expansion and satisfies a no-scale relation $g^{T\bar{T}} \del_T K \del_{\bar{T}} K=3$. As a consequence the $F$-term scalar potential vanishes identically, and the K\"ahler modulus is a flat direction. 
Crucially, the complex number $W_0$ is believed to be (discretely but finely) tunable. 

\begin{figure}
	\centering
	\includegraphics[width=10cm,keepaspectratio]{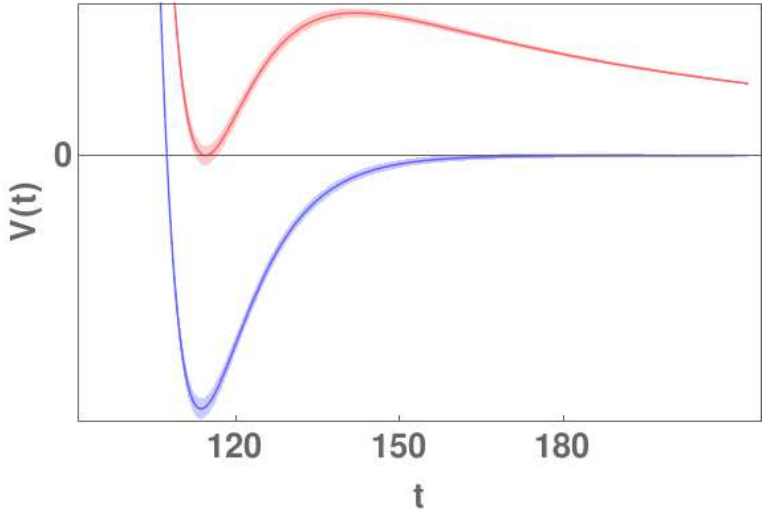}
	\caption{In blue: The supersymmetric KKLT potential for parameters $N A=1$, $\frac{2\pi}{N}=0.1$ and $W_0=-10^{-4}$. In red: The uplifted potential with appropriately fine tuned warp factor $a_0^2\sim |W_0|$. There is a dS minimum at $t\equiv \text{Re}(T)\approx 110$. The maximal possible uplift occurs with vacuum energy $V\sim +|V_{SUSY}|$. At larger values of $a_0^2$ there is no meta stable minimum anymore.}
	\label{fg:KKLTplot}
\end{figure}

KKLT have argued that the K\"ahler modulus $T$ can be stabilized as follows: Consider wrapping a stack of $N$ seven-branes around a representative four-cycle $\Sigma$. The low energy degrees of freedom that live on the stack organize into an $\mathcal{N}=1$ $SU(N)$ gauge multiplet and charged matter multiplets. If all charged matter can be given a large mass, the effective field theory is that of pure $SU(N)$ supersymmetric Yang-Mills (SYM) theory. At some high scale $\mu_0$ the holomorphic gauge coupling is set by the value of the K\"ahler modulus, $\tau_{YM}(\mu_0)=iT$. Pure SYM is known to undergo gaugino condensation and generate a non-perturbative contribution to the superpotential \cite{Veneziano:1982ah,Affleck:1983mk},
\begin{equation}
W_{np}(T)=N\Lambda^3(1+\mathcal{O}(\Lambda^3/M_P^3))=N A \exp(-2\pi T/N)+\mathcal{O}(e^{-2\cdot 2\pi T/N})\, .
\end{equation}
Here, $\Lambda^{3N}= \mu_0^{3N}e^{2\pi i \tau(\mu_0)}\equiv A^N e^{-2\pi T}$ is the holomorphic scale of the gauge theory. The expectation value of the gaugino bilinear is related to the non-perturbative superpotential via \cite{Veneziano:1982ah,Amati:1988ft}
\begin{equation}\label{eq:gauginocondensate_superpotential}
\langle \lambda\lambda \rangle\equiv \langle \Tr{\bar{\lambda}P_L\lambda}\rangle=-16\pi\del_T W_{np}(T)=32\pi^2 \Lambda^3\, .
\end{equation}
Summing up the classical flux superpotential and the gaugino condensation contribution we have
\begin{equation}\label{eq:KKLT_superpotential}
W\approx W_0+N A \exp(-2\pi T/N)\, .
\end{equation}
Now, the F-term condition $D_TW=0$ is solved by
\begin{equation}\label{eq:stabilizedT}
T\sim \frac{2\pi}{N}\log\left(-A/W_0\right)\, ,
\end{equation}
so the volume modulus is stabilized supersymmetrically (see figure \ref{fg:KKLTplot}). 
Clearly, for this construction to be under parametric control, it is necessary that $\text{Re}(T)\gg N$ because otherwise higher-order non-perturbative corrections become relevant \cite{Dine:1999dx}, and the gauge theory is not weakly coupled at the UV scale $\mu_0$. This is achieved by tuning $|W_0/A|\ll 1$. While this requirement renders the KKLT construction non-generic, it is believed that the huge number of choices of three-form flux quanta allows to tune the flux superpotential sufficiently finely \cite{Denef:2008wq}.

So far, the K\"ahler modulus is stabilized in a supersymmetric AdS vacuum. At the supersymmetric minimum the vacuum energy reads
\begin{equation}\label{eq:vacuumenergy_SUSYKKLT}
V_{SUSY}=-3e^{K}|W|^2\approx -\frac{3}{(T+\bar{T})^3}|W_0|^2\, .
\end{equation}
KKLT have proposed to deform this solution by adding SUSY breaking anti-branes at the bottom of so called warped throats (or Klebanov-Strassler throats) \cite{Klebanov:2000nc,Klebanov:2000hb}. These are strongly red-shifted regions in the compactification that are believed to be ubiquitous features of flux compactifications \cite{Giddings:2001yu,Hebecker:2006bn}. The red shift factor (warp factor) at the bottom of the throat is of order
\begin{equation}\label{eq:IRwarpfactor}
    a_0^2\sim \exp\left( -\frac{4\pi}{3}\frac{K}{g_s M}\right)\, ,
\end{equation}
where $(K,M)$ are a pair of integer flux numbers that thread the three-cycles of the throat \cite{Giddings:2001yu}. An anti-brane is naturally drawn to this region where it minimizes its scalar potential, given by \cite{Kachru:2003sx}
\begin{equation}\label{eq:KKLMMT}
    V_{\text{uplift}}(t)\sim \frac{a_0^4}{t^2}\, ,
\end{equation}
where $t\equiv \text{Re}(T)$. Moreover, a single anti-brane is believed to be stable if $M\gtrsim 12$ \cite{Kachru:2002gs,Almuhairi:2011ws,Polchinski:2015bea,Cohen-Maldonado:2015lyb,Bena:2018fqc}, and the $10d$ SUGRA solution is weakly curved in the throat region if $g_sM \gg 1$ \cite{Klebanov:2000hb}.  By tuning $a_0^4\sim |W_0|^2\ll 1$ one finds de Sitter vacua of the effective potential.\footnote{Recently it has been argued that the $D3$ brane tadpole forbids the existence of sufficiently strongly warped throats in toroidal setups \cite{Bena:2018fqc}. While this is true, in actual CY’s the tadpole bound becomes considerably weaker because of generically large amounts of negative induced $D3$ brane charge on seven brane stacks. We believe that very long throats are available.}

\section{KKLT: Gaugino condensates as classical sources}\label{sec:Gauginos}


In this section we would like to test the $10d$ gaugino condensation conjecture in the following simple way.\footnote{This test is too simple-minded to give strong evidence \textit{for} the conjecture (in contrast to e.g. \cite{Baumann:2010sx,Dymarsky:2010mf}). However, it shows very explicitly for what types of physical questions it \textit{cannot} hold.} We consider the classical $4d$ SUGRA Lagrangian of the KKLT theory as a function of the (constant) expectation value of the gaugino condensate that we insert by hand. We compare this ad-hoc quantity with the low energy effective scalar potential and find that they match to good approximation once the correct value of the gaugino condensate as given in eq. \eqref{eq:gauginocondensate_superpotential} is inserted. 

We start by considering the microscopic Lagrangian of 4d $\mathcal{N}=1$ pure $SU(N)$ Yang-Mills coupled to the K\"ahler modulus $T$, and the gravity multiplet. Our conventions are as in \cite{Freedman:2012zz} such that
\begin{equation}
e^{-1}\mathcal{L}_{gauge}=-\frac{1}{4}\text{Re}(f_{AB}(T))F_{\mu\nu}^AF^{\mu\nu\, B}+...\, ,
\end{equation}
with gauge kinetic function $f_{AB}=\frac{T}{4\pi}\delta_{AB}$. We take the K\"ahler and superpotential to be as in \eqref{eq:KKLT_Kahler_superpotential_classical}. It is useful to define the composite glueball field $S$ as
\begin{equation}
S\equiv \frac{\delta_{AB}\langle\bar{\lambda}^A P_L\lambda^B\rangle}{16\pi}\equiv\frac{\langle\lambda \lambda \rangle}{16\pi}\, .
\end{equation} 
Let us now focus on the gaugino mass term and quartic gaugino interactions,
\begin{align}
\mathcal{L}_{\lambda}=&-\frac{1}{16\pi}(T+\bar{T})^{-1/2}\overline{W_0}\Tr{\bar{\lambda}P_L \lambda}+h.c. -\frac{1}{48}\left(\frac{T+\bar{T}}{4\pi}\right)^2|\Tr{\bar{\lambda}P_L \lambda}|^2\, ,\nonumber\\
&+\frac{3}{64}\left(\frac{T+\bar{T}}{8\pi}\right)^2 \Tr{\bar{\lambda}\gamma_{\mu}\gamma_*\lambda}\Tr{\bar{\lambda}\gamma^{\mu}\gamma_*\lambda}\, .
\end{align}
We now ask what is the effective scalar potential $V(T,\bar{T},S,\bar{S})$ once we insert by hand a non-vanishing expectation value $S \neq 0$, assuming that expectation values for quartic terms factorize into the bilinear expectation values.\footnote{For evidence that this is actually the case, see e.g. \cite{Novikov:1985ic}.} Due to Lorentz invariance $\langle \Tr{\bar{\lambda}\gamma^{\mu}\gamma_* \lambda}\rangle=0$. Clearly, we obtain
\begin{equation}\label{eq:Vmicro}
V_{\text{microscopic}}(T,\bar{T},S,\bar{S})=\frac{(T+\bar{T})^2}{3}|S|^2+\frac{S\, \overline{W_0}+h.c.}{(T+\bar{T})^{1/2}}\, .
\end{equation}
Let us call this the \textit{microscopic potential}. It is not obvious that this expression is physically meaningful. Really, we should have used the fact that the gauge theory is gapped, integrated out all of its degrees of freedom and determined the effective scalar potential for the K\"ahler modulus. The effective superpotential below the mass scale of the gauge multiplet is given in eq. \eqref{eq:KKLT_superpotential}
so the scalar potential is given by the usual F-term scalar potential
\begin{align}\label{eq:Veffective}
V_{\text{effective}}(T,\bar{T})&=e^K\left(g^{T\bar{T}}|D_TW|^2-3|W|^2\right)\nonumber\\
&=\frac{1}{3(T+\bar{T})}\left(1+\frac{3}{2\pi \frac{\text{Re}(T)}{N}}\right)\left|2\pi Ae^{-2\pi T/N}\right|^2+\frac{2\pi A e^{-2\pi T/N}\overline{W_0}+h.c.}{(T+\bar{T})^2}\, .
\end{align}
This is the physically meaningful low energy effective potential. However, one notices immediately that in the limit of small t'Hooft coupling $\text{Re}(T)\gg N$, the low energy effective potential is actually the same as the microscopic potential upon plugging in the well known value for the gaugino condensate of eq. \eqref{eq:gauginocondensate_superpotential} (weighed by a normalization factor $e^{K/2}$)
\begin{equation}
S=2\pi e^{K/2}Ae^{-2\pi T/N}\, .
\end{equation}
So we see that to good approximation the scalar potential is given by (minus) the classical action evaluated at the correct value of the gaugino bilinear. The strong gauge dynamics essentially only freezes the gauge degrees of freedom and sets the expectation value of the gaugino condensate. Up to these effects, the classical action seems to approximate the low energy effective potential very well. Note that recently it has been understood how precisely this structure arises from a ten-dimensional treatment of gaugino condensation \cite{Hamada:2018qef,Kallosh:2019oxv}. Therefore we expect that the approximate equality between effective and microscopic potential holds also when the $4d$ theory is lifted to the full $8d$ gauge theory on the seven-brane stack embedded into the $10d$ bulk. However, at this point we would like to emphasize that
\begin{equation}
\del_{T}V_{\text{effective}}\approx \del_T V_{\text{microscopic}}+\frac{\del S(T)}{\del T}\del_S V_{\text{microscopic}}\neq \del_T V_{\text{microscopic}}\, ,
\end{equation}
so whenever derivatives of the scalar potential with respect to the K\"ahler modulus become relevant, special care is needed. In particular, this implies that all physical observables that are sensitive to derivatives of the scalar potential cannot be computed by treating gaugino condensates as classical source terms in a local $10d$ action. This is because the definition of $T$ is non-local from the $10d$ perspective, and $S$ varies exponentially with $T$. In the following section we will show that $V'(T)$ indeed plays a crucial part in ten dimensional tadpole cancellation requirements.

\section{The tadpole cancellation puzzle and its solution}\label{sec:tadpole_cancellation}


In this section we would like to first explain what the tadpole cancellation problem is and how we think it is resolved. We consider ten dimensional type IIB SUGRA with bosonic action
\begin{equation}
\label{IIBaction}
S_{IIB}=2\pi \int_{M_{10}}\left(*R_{10}-\frac{d\tau\wedge *d\overline{\tau}}{2(\text{Im}(\tau))^2}-\frac{G_3\wedge *\overline{G_3}}{2\text{Im}(\tau)}-\frac{F_5\wedge *F_5}{4}-\frac{iC_4\wedge G_3\wedge\overline{G_3}}{4\text{Im}(\tau)}\right)+S_{loc}\, ,
\end{equation}
with conventions as in \cite{Giddings:2001yu} but working in a system of units where $M_{10d}^8\equiv 1/\kappa^2_{10}=4\pi$. We consider compactifications down to four-dimensions with the isometries of $4d$ (anti-)de Sitter or Minkowski space. We may parametrize the $10d$ Einstein frame metric as,
\begin{equation}\label{eq:10D_4D6D_metric}
ds^2=G_{MN}dX^M dX^N=e^{2A}g^4_{\mu\nu}dx^{\mu}dx^{\nu}+g^6_{mn}dy^mdy^n\, .
\end{equation}
Here, $g^4_{\mu\nu}$ is the $4d$ metric which we take to be maximally symmetric, and $e^{2A}=e^{2A}(y)$ is the warp factor. The most general ansatz for the five-form field strength that is compatible with the $4d$ isometries is
\begin{equation}
F_5=(1+*)d\alpha\wedge \sqrt{-g^4}d^4x\, ,
\end{equation}
for some function $\alpha(y)$. The ten dimensional equations of motion imply that \cite{Giddings:2001yu,deAlwis:2003sn,Dasgupta:1999ss}
\begin{equation}
\tilde{\nabla}^2\Phi^-=R_{4D} +e^{2A}\left(\frac{|(*_6-i)G_3|^2}{4\text{Im}(\tau)}+e^{-8A}|\del \Phi^-|^2\right) + e^{2A}\frac{\Delta^{loc}}{2\pi} \, ,
\end{equation}
where $\Phi^-\equiv e^{4A}-\alpha$, $R_{4D}$ is the Ricci scalar of the metric $g^4$, and $\tilde{\nabla}^2$ is the $6d$ Laplacian of the auxiliary metric $\tilde{g}_{mn}\equiv e^{2A}g_{mn}^6$. Moreover, $\Delta^{loc}$ encodes the contributions from localized objects,
\begin{equation}\label{eq:Delta}
\Delta^{loc}=\frac{1}{4}\left(T^m_m-T^{\mu}_{\mu}\right)^{loc}-2\pi \rho_3^{loc}\, ,
\end{equation}
with $(T_{MN})^{loc}\equiv -\frac{2}{\sqrt{-G}}\frac{\delta S^{loc}}{\delta G^{MN}}$ denoting the stress-energy of localized sources, and $\rho_3^{loc}$ denoting the density of localized $D3$ brane charge. The tadpole puzzle arises because we may integrate the equation of motion over the internal manifold (with measure $\sqrt{\tilde{g}}$). The l.h. side must vanish since it is a total derivative so the r.h. side must vanish as well,
\begin{equation}\label{eq:NStadpole}
0\overset{!}{=}\int_{M_6}d^6y\sqrt{g_6}\left[e^{6A}R_{4D}+\underbrace{\frac{e^{8A}}{2\pi}\Delta^{loc}+\text{positive semi-definite}}_{\equiv \frac{e^{8A}}{2\pi}\Delta}\right]\, .
\end{equation}
It turns out that $\Delta^{loc}$ vanishes for a wide range of localized objects. In particular, all the ingredients considered in the GKP solutions do not contribute at all \cite{Giddings:2001yu}. This fact makes it difficult to find solutions with the isometries of four dimensional de Sitter space in the type IIB flux landscape, at least from the ten dimensional point of view. This is because such solutions must satisfy $R_{4D}>0$ so at least a single localized source must exist that gives a negative contribution to $\Delta^{loc}$. Such sources are simply hard to find. In particular anti $D3$ branes give a positive contribution. We will call this the tadpole problem.

This problem seems to be particularly severe for the KKLT mechanism. Let us assume that the $10d$ gaugino condensation conjecture is correct. In other words we assume that the $10d$ effects of the non-perturbative dynamics on the seven-brane stack can be described by inserting by hand an expectation value for the gaugino condensate and treating it as a classical source term. The four dimensional vacuum energy and thus the Ricci scalar of the KKLT vacuum is given by
\begin{equation}
R_{4D}\sim -|\langle\lambda \lambda\rangle|^2\, ,
\end{equation} 
where $\langle\lambda \lambda\rangle$ is the value of the gaugino bilinear of a seven-brane gauge theory wrapping some divisor $\Sigma$. Such vacua should lift to full $10d$ solutions so the gaugino condensate must enter the integrand of \eqref{eq:NStadpole} in a way such that the full tadpole is canceled, i.e. there should be a contribution to $\Delta^{loc}$ of the form
\begin{equation}\label{eq:Delta_gc}
\Delta^{loc}\supset \Delta_{\text{g.c.}}\sim +|\langle\lambda \lambda\rangle|^2\, .
\end{equation}
A complete and unambiguous evaluation of this effect has not been performed yet, for a recent discussion see \cite{Gautason:2018gln,Hamada:2018qef,Kallosh:2019oxv}. Of course, the expression \eqref{eq:Delta_gc} should only be valid when evaluated on the vacuum solution, where $W_0\sim \langle \lambda \lambda \rangle$. Away from the minimum one would expect it to be a combination of the gaugino condensate $\langle\lambda \lambda\rangle$ and the $(0,3)$ component of three-form fluxes $W_0$ that reduce to \eqref{eq:Delta_gc} when $\langle\lambda \lambda\rangle\sim W_0$, so really (as in eq. \eqref{eq:Vmicro})
\begin{equation}\label{eq:DeltaLoc_exp}
\Delta_{\text{g.c.}}\sim c_1 |\langle\lambda\lambda\rangle|^2+c_2 \text{Re}\left(\overline{W_0}\langle \lambda\lambda \rangle\right)\, ,
\end{equation}
for some coefficients $c_{1,2}$ that have at most a power-law dependence on the overall volume.
 
If this were the only contribution from the seven-brane gauge theory we would be left with a problem. First, due to the eight powers of the warp factor that multiply the integrand of the tadpole, warped SUSY breaking objects such as anti-branes give negligible contributions via their stress energy tensor. Second, according to the $4D$ description of KKLT, the $AdS$ minimum can be perturbed to actually become a Minkowski solution or even a de Sitter solution, \textit{without} significantly affecting the values of either $W_0$ or the gaugino condensate. For this to be consistent the contribution to the tadpole from the seven-brane sector would have to actually change in sign. It thus seems that the tadpole cannot be cancelled on the uplifted solution unless there is an almost cancellation between the two terms in eq. \eqref{eq:DeltaLoc_exp} when evaluated in the vacuum. So far, no structure on the seven-brane Lagrangian has been found that would allow this to occur. We will now argue that indeed away from the supersymmetric minimum there is a generically dominant further contribution to the tadpole which will resolve this puzzle. We interpret this contribution as the stress caused by the \textit{restoring force} that acts on the volume modulus once it is displaced from its supersymmetric minimum, and is likely not captured by the \textit{local} $10D$ action.

Let us first be completely open minded to what is the form of the $D7$-brane action that perturbs the GKP background. We only write
\begin{equation}
S^{D7}[G]=\int d^4 x\int d^6y \sqrt{-G}\tilde{\mathcal{S}}[g_6](x,y)\, ,
\end{equation}
where $\tilde{\mathcal{S}}$ is some yet unspecified space-time dependent functional of the internal metric. We now restrict ourselves to trivial warping. The stress energy tensor of $S^{D7}[G]$, evaluated on the background given in eq. \eqref{eq:10D_4D6D_metric} reads 
\begin{align}
T_{\mu\nu}^{D7}&=g_{\mu\nu}^4 \tilde{\mathcal{S}}[g_6](y)\, ,\\
T_{mn}^{D7}&=-\frac{2}{\sqrt{g_6}}\frac{\delta \mathcal{S}[g_6]}{\delta g_6^{mn}}\, ,\quad \text{with}\quad \mathcal{S}[g_6]\equiv \int d^6y \sqrt{g_6}\tilde{\mathcal{S}}[g_6]\, .
\end{align}
We are now ready to evaluate the contribution to the tadpole coming from $S^{D7}[g]$.\footnote{For simplicity we will neglect its dependence on $C_4$, and thus any contributions to $\rho_{3}^{loc}$. The background induced $D3$ brane charge will cancel against the $7$-brane tension at order $\alpha'^2$ \cite{Giddings:2001yu}, and contributions from the non-perturbative stabilization of the $C_4$ axion will vanish as long as the axion is not displaced from its minimum. We will not consider sources that displace the axion from its SUSY minimum.} We may expand the (inverse) internal metric $g^{mn}$ in a complete set of symmetric two-tensors $\{S_i^{mn}\}$,
\begin{equation}
g^{mn}=\sum_i a_i S^{mn}_i\, ,
\end{equation}
with Fourier coefficients $a_i$. These can for instance be taken as eigen-functions of the linearized Einstein equations which corresponds to the usual KK-mode expansion. We take the completeness relation to be
\begin{equation}
\int_{M_6}d^6y \sqrt{g_0}S_i^{mn}S_{j\, mn}=\delta_{ij}\, ,
\end{equation}
where $g_0$ is a solution to the $10d$ equations of motion (before the inclusion of $S_{D7}$). Hence, the inverse relation is
\begin{equation}
a_i=\int d^6 y \sqrt{g_0}\, S_{i\, mn}g^{mn}\, .
\end{equation}
We can now vary the functional $\mathcal{S}$ with respect to the Fourier coefficient $a_i$,
\begin{align}
\frac{\del \mathcal{S}[g]}{\del a_i}=&\int d^6y \frac{\del g^{mn}(y)}{\del a_i}\frac{\delta \mathcal{S}[g]}{\delta g^{mn}(y)}\\
=&\int d^6y \sqrt{g}\,  S_i^{mn}\frac{1}{\sqrt{g}}\frac{\delta \mathcal{S}[g]}{\delta g^{mn}}=-\frac{1}{2}\int d^6 y \sqrt{g}S^{mn}_i T^{D7}_{mn}\,  ,
\end{align}
where in the last equality we have implemented the definition of the stress-energy tensor. We may now choose our complete set of tensors ${S_i^{mn}}$ such that $S_0^{mn}=g^{mn}_0$, so that $a_0\equiv \lambda$ is the Fourier coefficient corresponding to the overall volume modulus. Moreover, we may consider a one-parameter family of solutions $g^{mn}=\lambda g^{mn}_0$. Then,
\begin{equation}\label{eq:delSdellambda}
-\frac{1}{2}\lambda\frac{\del \mathcal{S}[g]}{\del \lambda}=\int d^6 y \sqrt{g}\frac{1}{4}T^m_{m}\, .
\end{equation}
Furthermore, it is easy to see that
\begin{equation}
\mathcal{S}[g]=\int d^6 y\sqrt{g}\frac{1}{4}T^{\mu}_{\mu}\, ,
\end{equation}
so there is a contribution to the tadpole,
\begin{equation}
\int d^6 y\sqrt{g}\Delta^{loc,D7}=-\left(\frac{1}{2}\lambda\frac{\del}{\del \lambda}-1\right)\mathcal{S}[g(\lambda)]\, .
\end{equation}
Since we consider static solutions, we should interpret $\mathcal{S}[g]$ as a contribution to the $4d$ scalar potential from the seven-brane stack. Concretely, let us define the overall volume of the CY as
\begin{equation}
t^{3/2}\equiv \int d^6y \sqrt{g_6}\, ,
\end{equation}
and let $t_0$ be its value for $g=g_0$ (so $\lambda=(t_0/t)^{1/2}$). Then, the four dimensional Einstein frame metric is
\begin{equation}
g^{E}_{\mu\nu}=(t/t_0)^{3/2}g^4_{\mu\nu}\, ,
\end{equation}
and the action $S^{D7}$ reads
\begin{equation}
S^{D7}[G]=\int d^4x \sqrt{-g^E}\left(\frac{t_0}{t}\right)^3\mathcal{S}[g]\, .
\end{equation}
Therefore, the four dimensional scalar potential is
\begin{equation}
V_{D7}(t)\equiv -\left(\frac{t_0}{t}\right)^{3}\mathcal{S}[g(t)]\, .
\end{equation}
We may then write the tadpole bound in the simple form
\begin{equation}\label{eq:tadpoleconstraint_final}
\frac{1}{4}R^{E}M_P^2-V_{D7}(t)-\frac{1}{2}t\del_t V_{D7}(t)=0\, .
\end{equation}
$R^E$ denotes the $4d$ Ricci scalar of the Einstein frame metric $g^E$. We have used that $M_P^2=4\pi t_0^{3/2}$, and that $R^E=(t_0/t)^{3/2}R_{4D}$, and finally that there are no further contributions to the tadpole. 

As a consequence the tadpole is canceled whenever only the seven-branes contribute to the scalar potential, and if $V_{D7}(t)$ has a minimum. This is because (on a maximally symmetric background) the $4d$ Einstein equations reduce to $R^EM_P^2=4V_{D7}(t)$, while the equation of motion for the volume modulus are solved when $V_{D7}'(t)=0$.

From this expression we learn the following: Assuming the consistency of the supersymmetric KKLT construction, the tadpole is canceled at the supersymmetric minimum and the only non-vanishing contributions to the tadpole come from the $4D$ Ricci scalar and the seven-brane scalar potential. If the approximate equivalence between the microscopic \eqref{eq:Vmicro} and the effective scalar potential \eqref{eq:Veffective} lifts also to ten dimensions there is good reason to believe that the supersymmetric four dimensional KKLT vacua can be lifted to solutions of the ten dimensional equations of motion, treating the gaugino bilinear as a classical source term. This is the approach followed in \cite{Baumann:2010sx,Dymarsky:2010mf,Moritz:2017xto,Gautason:2018gln}. However, once the SUSY minimum is left, there is a qualitatively new contribution to the tadpole which is proportional to $V_{D7}'(t)$. This term is interpreted as the \textit{restoring force} that non-perturbative seven-brane effects exert on the volume modulus. As we saw in section \ref{sec:Gauginos}, such terms receive dominant contributions from derivatives of $e^{-2\pi T/N}$ with respect to $T$. We find it very unlikely that such contributions to the tadpole can be encoded in a local $10D$ action. Rather, we believe that the ability to describe KKLT vacua by inserting a local $10D$ action is limited to cases without stress from restoring forces.

We would now like to apply this result to warped uplifts. Strictly speaking we have not considered warping in the above discussion. But it seems very reasonable to us that the existence of strongly warped regions does not alter the stress energy tensor of the seven-brane stack. This is because we assume that it wraps a four-cycle in the essentially unwarped bulk.  A warped uplift is characterized by sub-leading contributions to the integrand of the tadpole via its stress energy tensor. This is because such contributions are suppressed by eight powers of the warp factor in the tadpole constraint \eqref{eq:NStadpole}. However, an uplift (if sufficiently small) will affect the solution in two ways:
\begin{enumerate}\label{list:uplift-effects}
	\item It will pull the volume modulus toward larger values, $t-t_{SUSY}>0$.
	\item It will raise the vacuum energy.
\end{enumerate}
This must happen in such a way that the $4d$ Einstein equations as well as the equations of motion of the $t$ modulus are solved, i.e.
\begin{equation}
\frac{1}{4}R^EM_P^2=V_{D7}(t)+V_{uplift}(t)\, ,\quad 0=V_{uplift}'(t)+V_{D7}'(t)\, .
\end{equation}
Plugging this into the tadpole constraint of eq. \eqref{eq:tadpoleconstraint_final} we find that it is canceled provided that also
\begin{equation}
V'_{uplift}(t)=-\frac{2}{t}V_{uplift}(t)\, .
\end{equation}
But this equation is actually satisfied if the warped uplift potential is well approximated by the classical expression of \eqref{eq:KKLMMT}.

Thus, we have shown that any seven-brane action that reproduces the KKLT scalar potential upon dimensional reduction to four dimensions will automatically ensure tadpole cancellation upon uplifting by warped SUSY breaking sources. It is natural to ask, what actually \textit{is} the precise form of this action. In general it is not possible to uniquely infer the higher dimensional scalar potential from the lower dimensional one, but there is one very natural candidate action. The $10d$ action proposed in ref. \cite{Hamada:2018qef,Kallosh:2019oxv} gives rise to the microscopic $4d$ potential of eq. \eqref{eq:Vmicro} upon dimensional reduction as shown in \cite{Hamada:2018qef}. The $4d$ microscopic potential is approximately equivalent to the true effective potential (as we saw in section \ref{sec:Gauginos}). Thus, by inserting,
\begin{equation}
\langle \lambda \lambda \rangle \propto e^{-2\pi T[g,C_4]/N}\, ,\quad \text{with}\quad  T[g,C_4]\equiv \int_{\Sigma}\left(d^4y\sqrt{g_{\Sigma}}+i C_4\right)\, ,
\end{equation}
in the $10d$ action \cite{Hamada:2018qef,Kallosh:2019oxv}, one obtains a (non-local) action $S_{D7}[g,C_4]$ that generates the $4d$ KKLT scalar potential, the correct potential for $D3$ brane position moduli as in \cite{Baumann:2010sx}, and is consistent with tadpole cancellation upon uplifting by warped sources.

The above reasoning is orthogonal to the question whether or not it is actually possible to generate sufficiently long warped throats and decouple the stabilization sector from the uplift sector efficiently. We have simply assumed that this can be done. Rather it shows that \textit{if} these requirements are assumed to be met in a controlled manner, $10d$ tadpole cancellation does not indicate an inconsistency of the assumptions that were made. We will comment on difficulties to actually realize this in section \ref{sec:warped_uplifts}.

As a side remark we note that the above gives further justification to recent approaches to tadpole cancellation in $10d$ calculations where the tadpole is canceled via the by-hand addition of extra source terms \cite{Kim:2018vgz}. These terms should really be interpreted as a dynamically adjusting restoring force against decompactification.

\vspace{0.5cm}

\noindent\textbf{Note added:} Two other papers addressing the same problem were uploaded to the arXiv simultaneously with 
ours: \cite{Hamada:2019ack,Gautason:2019jwq}. Ref. \cite{Hamada:2019ack} reaches the same conclusions regarding the form of the stress energy tensor from gaugino condensation, and thus the compatibility of the KKLT model with ten dimensional tadpole cancellation constraints. However, ref. \cite{Gautason:2019jwq} comes to the opposite conclusion by treating the expectation value of the gaugino condensate as a purely classical source term. This means neglecting any terms proportional to $\del_T \langle \lambda\lambda \rangle$. We have argued that the stress energy tensor from the restoring force of the volume stabilization mechanism receives dominant contributions from such terms. In principle it might be consistent to treat the gaugino condensate as an independent field, but one would then need to start with an off-shell action of $ \langle \lambda\lambda \rangle$ and $T$ which involves the Veneziano-Yankielowicz scalar potential \cite{Veneziano:1982ah} that stabilizes $\langle \lambda\lambda \rangle$ in terms of $T$. We think that including this potential in the seven-brane Lagrangian would resolve the tension between our results and those of ref. \cite{Gautason:2019jwq}.

\section{How small can a warped uplift be? }\label{sec:warped_uplifts}


In this section we will comment on the question if sufficiently long warped throats exist so that the KKLT effective field theory can be realized with appropriate values of parameters. By appropriate we mean that K\"ahler moduli stabilization occurs at sufficiently large volume, and the uplift potential is sufficiently small to prevent destabilization. In short, \textit{are parametrically small warped uplifts part of the landscape, or are they in the swampland}? Before we start we define small/large uplifts as follows.
\begin{align}\label{def:small/large_uplift}
	 &\textit{A small uplift is one that does not destabilize any of the moduli.}\nonumber\\
	&\textit{A large uplift is one that is not small.}
\end{align}
We will refer to a parametrically small uplift as one with parametrically negligible backreaction on all moduli. The moduli that will be relevant (i.e. the lightest ones) are the K\"ahler moduli which we assume to be stabilized non-perturbatively as in KKLT. 

Throughout this section we assume that both the flux number $W_0$ in the KKLT superpotential \eqref{eq:KKLT_superpotential} as well as the infra-red warp factor $a_0^2$ of the KS throat can be tuned arbitrarily well for all practical purposes. Thus, we will ignore the fact that the possibly finite number of flux vacua and $D3$-brane charge cancellation limit the extend to which this can actually be done \cite{Kachru:2003aw,Denef:2008wq}. Rather we will ask only for geometrical consistency of the setup once the volume modulus is stabilized via KKLT and the warp factor is small enough to prevent decompactification. We will see that this surprisingly hard to achieve.\footnote{Note that the results of \cite{Baumann:2010sx} show that in KKLT setups (at fixed value of $|W_0|$) indefinitely small uplifts are beyond the regime where the throat is well approximated by the KS solution. This is due to relevant perturbations of the KS gauge theory that are activated by gaugino condensation in the bulk CY and grow toward the infrared. However, parametrically small uplifts in the sense of \eqref{def:small/large_uplift} are not straightforwardly excluded by this.}

Let us briefly recall what are the qualitatively different regimes of values that the overall volume modulus $T$ can take \cite{Giddings:2001yu,deAlwis:2003sn,Giddings:2005ff,Shiu:2008ry} (see our appendix \ref{app:volume_modulus} for a summary). The obvious regime is that of very large values where a change in the modulus amounts to an overall rescaling of the volume of the compactification. Whenever at least one complex structure modulus $z$ is stabilized near a conifold singularity in complex structure moduli space, $|z|\ll 1$, this regime corresponds to \cite{Giddings:2005ff}
\begin{equation}
\text{Re}(T)\gg N_{D3}|z|^{-4/3}\, ,
\end{equation} 
where $N_{D3}$ is the total $D3$ brane charge stored in fluxes that thread the $\mathcal{A}$ and $\mathcal{B}$ cycle of the throat (and possibly in mobile $D3$ branes). This regime (called the dilute flux regime) is characterized by negligible backreaction via fluxes.

The warped throat regime occurs for values
\begin{equation}
N_{D3}|z|^{-4/3}\gg  \text{Re}(T) \gg N_{D3}\, .
\end{equation}
For values in this range, at least one warped throat forms with localized significant backreaction via fluxes that drive the non-trivial warping. The warped throat can be thought of as an object of size $R^4_{\text{throat}}\sim N_{D3}$ glued into a much larger bulk CY that remains unaffected by flux backreaction. Changing the value of $\text{Re}(T)$ essentially only rescales the bulk CY but leaves the throat unchanged. While flux backreaction is significant, it is controlled by the KS solution \cite{Klebanov:2000hb} which is smoothly glued into the bulk CY (see figure \ref{fg:throat_fit1}).

\begin{figure}
	\centering
	\begin{tabular}{ c c}
		\includegraphics[width=7cm,keepaspectratio]{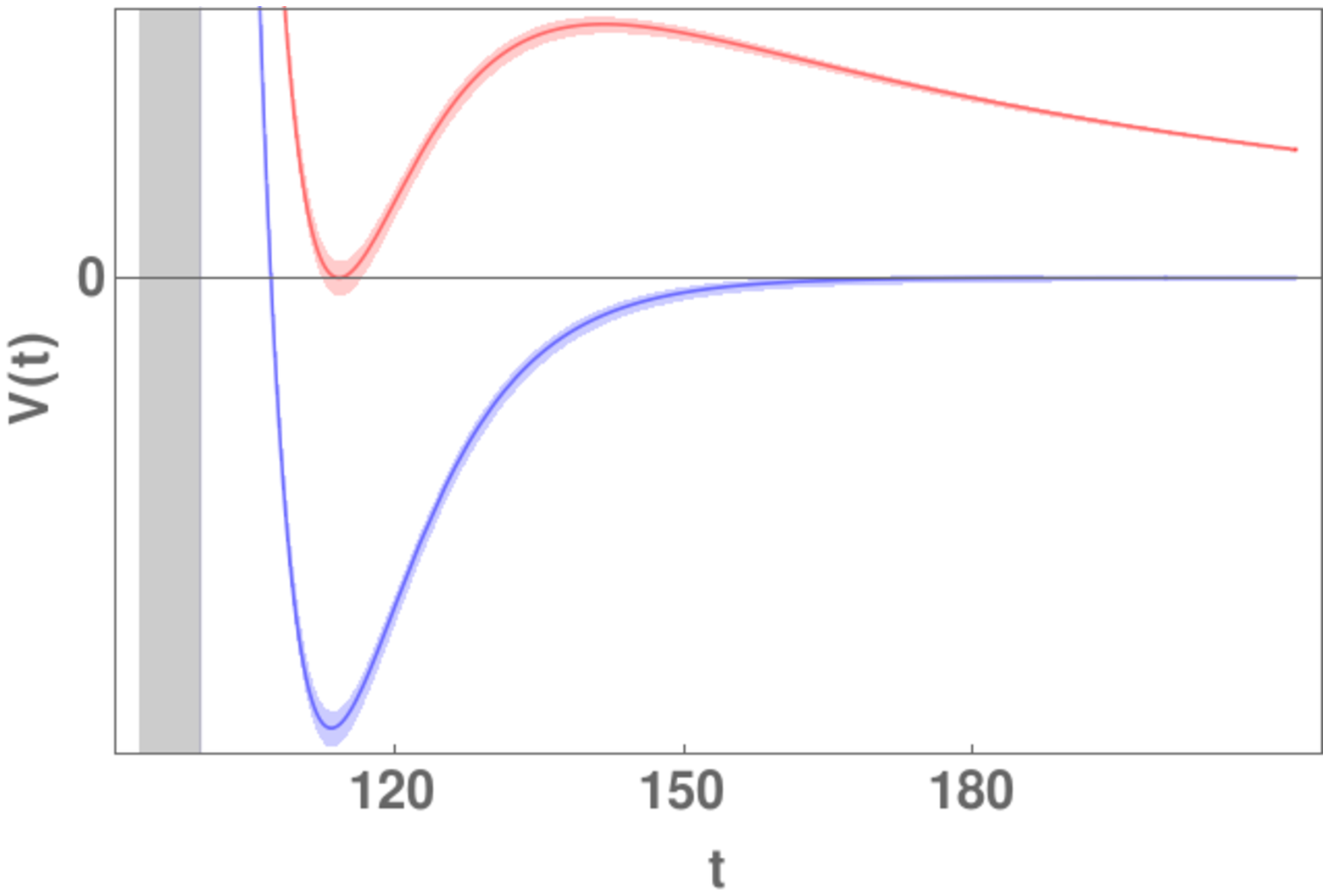} &
		\includegraphics[width=7cm,keepaspectratio]{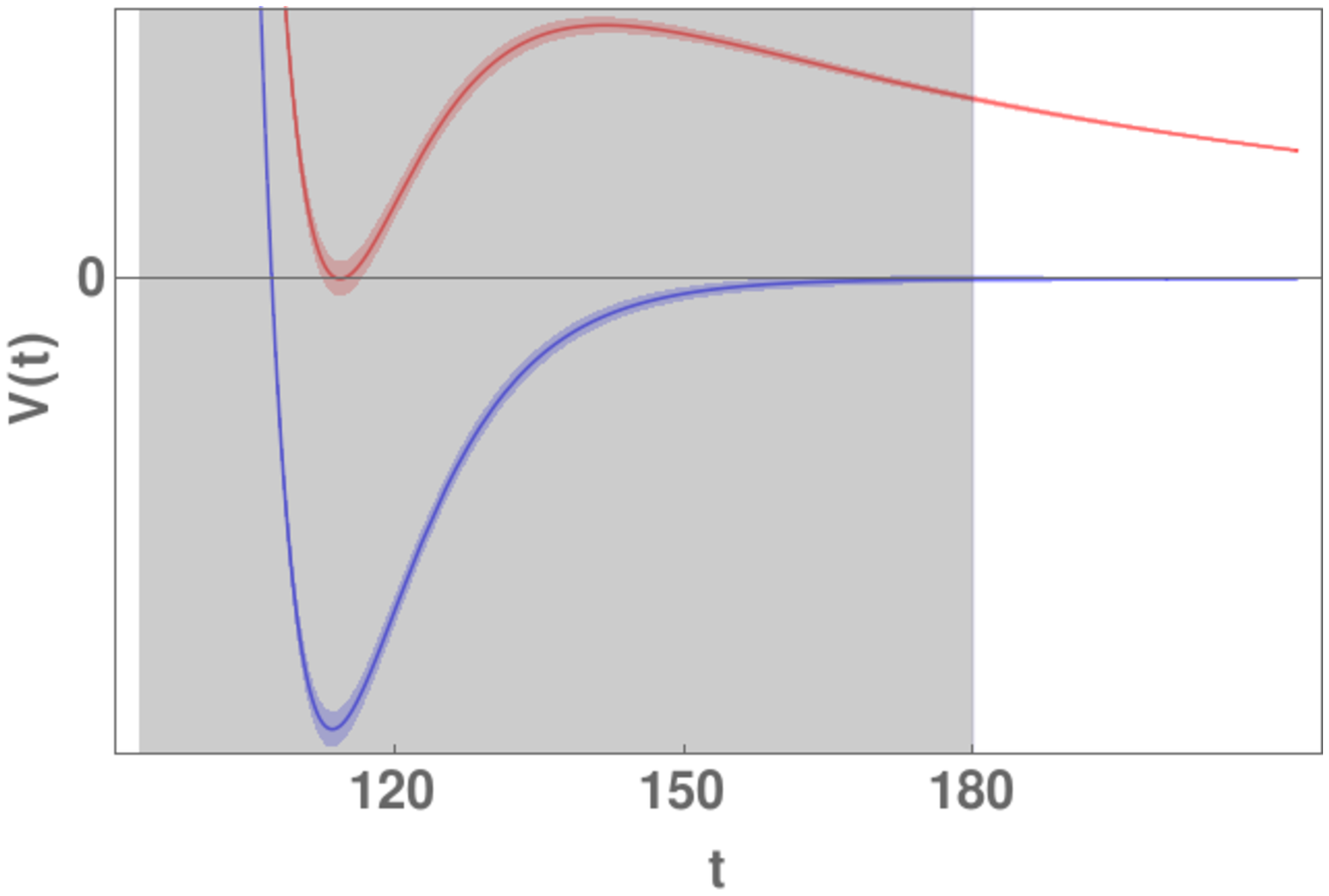}
	\end{tabular}
	\caption{We plot the KKLT uplift as in figure \ref{fg:KKLTplot}, but with a shaded area that marks the region $\text{Re}(T)< N_{D3}$ where the $10d$ geometrical setup is beyond control. Left: If (say) $N_{D3}=100$ can be realized simultaneously with the parameter choices of figure \ref{fg:KKLTplot} the uplifted vacuum lies in a controlled region and the uplift can be trusted. Right: If $N_{D3}\gg 100$ is required, the vacuum region cannot be trusted. We find that the scenario on the left is hard or impossible to realize, while generically we are forced into the scenario displayed on the right.}
	\label{fg:KKLTplot2}
\end{figure}

This regime ends once we set $\text{Re}(T)\sim N_{D3}$. At this point flux backreaction becomes significant throughout the CY. Moreover, it is not clear to us what is the physical meaning of going beyond this. At least the ability to describe the physical setting via semi classical GKP solutions of $10d$ SUGRA starts to break down beyond this point in field space. We will constrain KKLT de Sitter uplifts by demanding that the point in field space where (supersymmetric) K\"ahler moduli stabilization occurs must not lie in the uncontrolled regime $\text{Re}(T)\lesssim N_{D3}$ (see figure \ref{fg:KKLTplot2}). 
We will find that this constraint can generically not be met simultaneously with the constraint that the warped uplift does not trigger decompactification.

We would like to emphasize that the arguments that follow are very similar to (and were inspired by) the ones used in ref. \cite{Freivogel:2008wm,McAllister:2008hb}. In particular, in \cite{Freivogel:2008wm} it was argued that $W_0$ has to be tuned extremely small in order for the KKLT construction to be able to work. However, we will argue that small uplifts are severly constrained even if $W_0$ can be tuned arbitrarily small. 

\subsection{Single modulus KKLT}
We consider the setup of section \ref{sec:KKLTsummary}, i.e. a type IIB Calabi-Yau orientifold with only a single K\"ahler modulus $T$ and with all complex structure moduli integrated out consistently.

Let us now see if we can make the warped uplift small in the sense defined in \eqref{def:small/large_uplift}. In order to suppress backreaction on $T$ we need that $V_{uplift}\lesssim |V_{SUSY}|$ \cite{Kachru:2003aw}. Comparing with eq. \eqref{eq:vacuumenergy_SUSYKKLT} and eq. \eqref{eq:KKLMMT} we see that we need $\alpha \equiv \frac{a_0^4}{|W_0|^2} \lesssim 1$, and as always $|W_0|^2\ll 1$. It is useful to define an uplift parameter $U$ as follows,
\begin{equation}
U\equiv \frac{\log(|V_{SUSY}|^{-1})}{\log{(|V_{uplift}|^{-1})}}\approx
	 \frac{\log |W_0|^{-2}}{\log a_0^{-4}}=1+\frac{\log(\alpha)}{\log(a_0^{-4})}\, ,
\end{equation}
in units $M_P=1$.
The uplift is small when $U \lesssim 1$. The interesting regime where an uplift to de Sitter space might take place corresponds to taking $\alpha =\mathcal{O}(1)$, so $|U - 1|\ll 1$. Once $U-1=\mathcal{O}(1)$ the uplift becomes \textit{exponentially} uncontrollable, i.e. 
\begin{equation}
    \alpha =\left(|W_0|^{-2}\right)^{\frac{U-1}{U}}\gg 1\, .
\end{equation}
The results of ref. \cite{Baumann:2010sx} imply that for control of relevant throat perturbations one needs $U \geq 3/4$, but only slightly bigger values allow to keep those perturbations under parametric control. We will now explain why only parametrically large values of $U$ might be possible in a geometrically consistent setup.

\begin{figure}
	\centering
	\includegraphics[width=11cm,keepaspectratio]{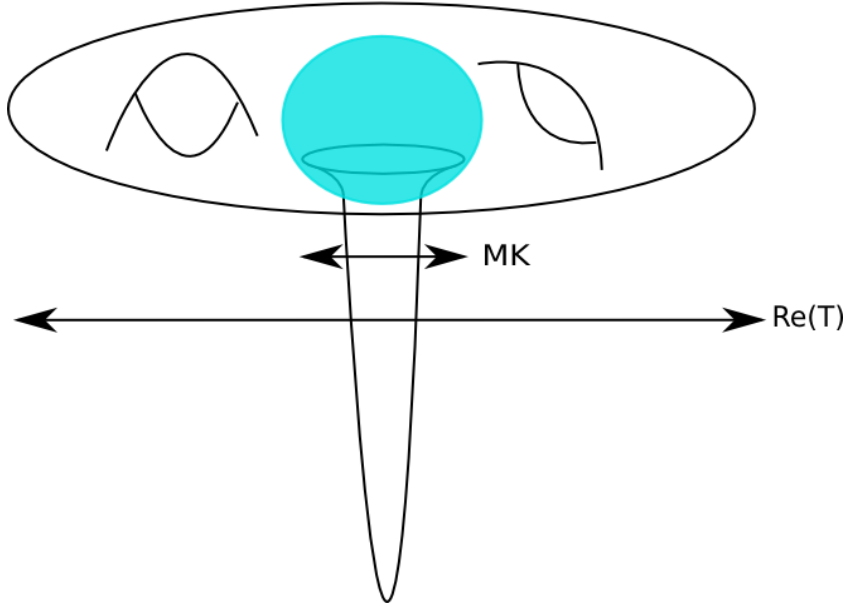}
	\caption{We depict a cartoon of a CY that contains a warped throat region. For geometrical consistency the throat region must be smaller than the overall size of the CY.}
	\label{fg:throat_fit1}
\end{figure}

This constraint comes from the following requirements. The throat that carries the SUSY breaking source has to be sufficiently small in circumference that it can fit into the bulk Calabi-Yau so that it can be well separated from the seven-brane stack that is responsible for K\"ahler moduli stabilization (see Figure \ref{fg:throat_fit1}). 
As recalled in the introduction of this section it is well-known that the (Einstein frame) size of the throat is set by its $D3$-brane charge $N_{D3}$ as
\begin{equation}
R_{throat}^4\sim N_{D3}=MK\, ,
\end{equation}
where $M$ and $K$ are the RR and NS-NS flux quanta that stabilize the throat. Thus we should require that
\begin{equation}\label{eq:throatfits}
\text{Re}(T)\gg MK\, ,
\end{equation}
for validity of the $10d$ geometric picture.

\begin{figure}
\centering
\includegraphics[scale=0.5]{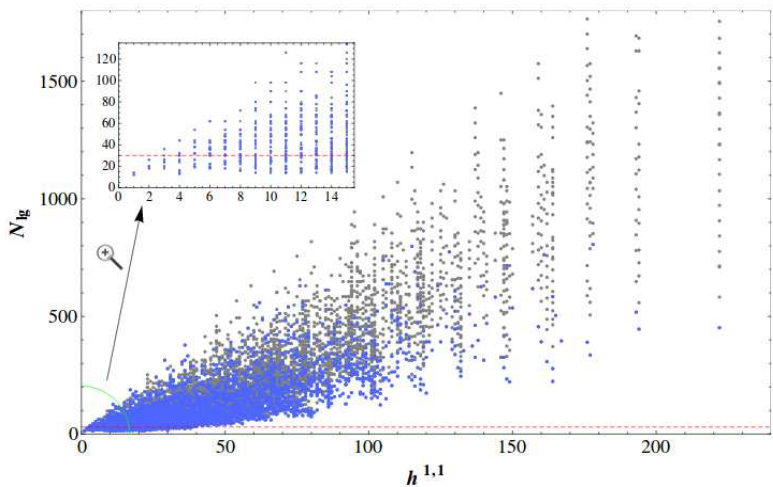}
\caption{For a subset of the Kreuzer-Skarke database \cite{Kreuzer:2000xy}, the maximal possible gauge group rank in perturbative type IIB string theory grows approximately linearly with $h^{1,1}$. This argument and the above figure are taken from \cite{Louis:2012nb}.
}
\label{fg:h11vsNgfig}
\end{figure}

However, at the KKLT minimum the value of $\text{Re}(T)$ is set by $W_0$ according to eq. \eqref{eq:stabilizedT} while the IR warp factor is set by eq. \eqref{eq:IRwarpfactor}. Plugging these formula into the requirement of eq. \eqref{eq:throatfits} we obtain
\begin{equation}
N\overset{!}{\gg} \frac{MK}{\log |W_0|^{-1}}\sim \frac{MK}{U \log a_0^{-4}}\sim  \frac{g_sM^2}{U}\, .
\end{equation}
This can be turned into a \textit{lower} bound on the uplift parameter $U$,
\begin{equation}
U \gg (g_sM)^2 g_{s}^{-1} \frac{1}{N}\, .
\end{equation}
Clearly we cannot make $U$ arbitrarily small. But, in fact it is not clear to us if it can even be smaller or equal to one as would be required for a controlled uplift. The size of the IR end of the throat as measured in string units is $g_sM$, so we need $g_sM\gg 1$ for control of the $\alpha'$ expansion, as well as $g_s\ll 1$ for control of the string loop expansion.
So in order to get $U\approx 1$ or smaller, the rank of the seven-brane gauge group really has to be parametrically large,
\begin{equation}\label{eq:bound_on_N}
N\gg (g_sM)^2 g_s^{-1}\gg 1\, .
\end{equation}
There is at least some indication that it is hard to make $N$ arbitrarily large while maintaining $h^{1,1}=1$. Indeed one of the swampland criteria states that in an EFT coupled to gravity the number of massless fields cannot be arbitrarily large \cite{Brennan:2017rbf}, and in this case an arbitrarily large $N$ would imply an arbitrarily large number of gluons. This generic swampland criterium can be checked explicitly in some setups. For example, in \cite{Louis:2012nb} the relation between $h^{1,1}$ and the largest possible $N$ was investigated numerically using a subset of the Kreuzer-Skarke database. For $h^{1,1}=1$, the largest possible $N$ was found to be $\mathcal{O}(10)$, while more generally a linear bound of the form
\begin{equation}\label{eq:NvsH11}
N\leq \mathcal{O}(10)h^{1,1}\, ,
\end{equation}
was found to be obeyed within the limited example set, see figure \ref{fg:h11vsNgfig}. Of course, it could be true that much larger values of $N$ exist for $h^{1,1}=1$ that are not contained in the example set. This possibility forms a potential loophole. We conclude the following.  
\begin{center}
	\textit{In single modulus KKLT with $N=\mathcal{O}(1)$ and $A=\mathcal{O}(1)$, whenever the supersymmetric starting point lies in a regime of parametric control of the supergravity theory, $\text{Re}(T)\gg MK$, all warped uplifts will destabilize the K\"ahler modulus.}
\end{center}
Assuming that the empirical relation of eq. \eqref{eq:NvsH11} holds, the loophole $N\gg 1$ is closed. The obvious remaining loophole is to take $h^{1,1}\gg 1$ which is anyway satisfied by generic CYs, and/or assuming $|A|\gg 1$. We will comment on these options momentarily.

\subsection{Multi modulus KKLT}
The arguments of the previous subsection indicate that it is impossible to realize parametrically small uplifts in the case in which the CY orientifold has just one K\"ahler modulus, i.e. whenever $h^{1,1}_+=1$. One could then try to evade this reasoning, by considering a case with many K\"ahler moduli. So, let us now consider the potential loophole $h^{1,1}_+\gg 1$.

We would like to emphasize immediately that most of the conclusions we draw in this section are based on assumptions about the geometry of CY manifolds which we believe to hold \textit{generically}. As such we cannot exclude that non-generic CYs exist for which our discussion does not apply. We will comment on this possibility at the end of this section.

Let us consider a number of K\"ahler moduli $\{T_1,...,T_{h^{1,1}_+}\}$, and KKLT type superpotential
\begin{equation}
W=W_0+\sum_{i=1}^{n}N_i A_i \exp\left(-\frac{2\pi}{N_i} \sum_{\alpha=1}^{h^{1,1}_+}k^{\alpha}_iT_{\alpha}\right)\, ,
\end{equation}
with some integer-valued $n\times h^{1,1}_+$ charge matrix $k$. For the K\"ahler potential in the multi-modulus case we refer the reader to ref. \cite{Grimm:2004uq}. Here, we have assumed that there are $n$ superpotential terms that are generated via various confining gauge theories ($N_i>1$) or euclidean D3 brane instantons ($N_i=1$).

\begin{figure}
	\centering
	\includegraphics[width=11cm,keepaspectratio]{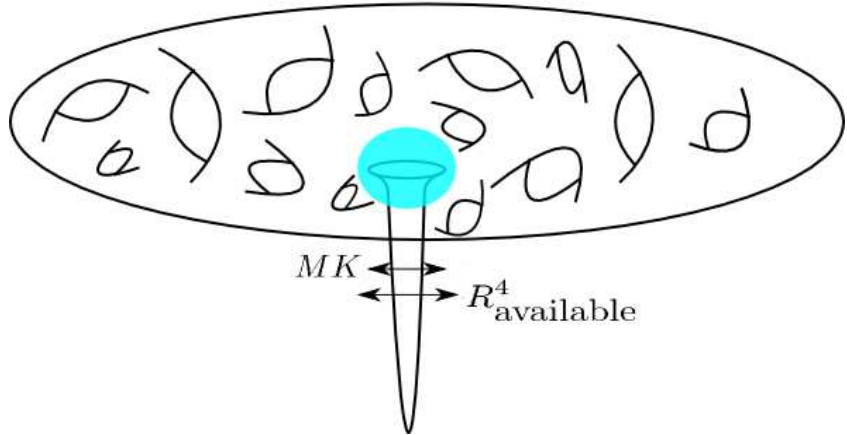}
	\caption{We draw a cartoon of a more complicated CY with $h^{1,1}\gg 1$. We expect that the rich topological structure of the manifold leaves us less room to place warped throats in comparison to a simpler CY with $h^{1,1}=1$ with the same overall volume (compare with figure \ref{fg:throat_fit1}).}
	\label{fg:throat_fit2}
\end{figure}

The F-term equations read
\begin{equation}
D_{T_{\alpha}}W=-\sum_{i=1}^{n}2\pi k_i^{\alpha} A_i \exp\left(-\frac{2\pi}{N_i} \sum_{\beta=1}^{h^{1,1}_+}k^{\beta}_iT_{\beta}\right)-\frac{2v^{\alpha}}{\mathcal{V}}W\, ,
\end{equation}
where $\mathcal{V}$ is the overall volume, and $v^{\alpha}$ is the volume of the two-cycle dual to the four-cycle $\Sigma^{\alpha}$. We expect the generic KKLT type solutions to satisfy,
\begin{equation}
\sum_{\alpha=1}^{h^{1,1}_+}k^{\alpha}_i\text{Re}(T_{\alpha})\sim \frac{N_i}{2\pi}\log(|W_0|^{-1})\, \quad \forall \, i ,
\end{equation}
so that generically the four-cycle volumes are again bounded as, 
\begin{equation}
\text{Re}(T_{\alpha})\lesssim \frac{N_{max}}{2\pi}\log(|W_0|^{-1})\, ,
\end{equation}
where $N_{max}$ is the maximal available dual Coxeter number. So far, the story is very similar to the case $h^{1,1}_+=1$, except that we expect to have the freedom to take $N_{max}\gg 1$.  Again we need to require that the throat fits into the bulk Calabi-Yau,
\begin{equation}
MK< \mathcal{V}^{2/3}\, .
\end{equation}
Now, we find it reasonable to demand something stronger: The throat must fit into a region in the bulk Calabi-Yau that is well approximated by a conifold region (or more generally some cone over a Sasaki-Einstein base). As a consequence we should really require that zooming into such a region, all the topological structure that the Calabi-Yau possesses should become invisible. This is because we would like to isolate the non-perturbative stabilization sector associated with \textit{each} $4$-cycle from the uplift sector associated to the throat (see Figure \ref{fg:throat_fit2}). 
At large $h^{1,1}_+$ it is natural to expect that in generic CY manifolds the amount of such freely available volume where warped throats can fit scales with the overall volume of the Calabi-Yau, but also that it decreases monotonically with $h^{1,1}$. For definiteness let us parameterize this expectation as
\begin{equation}
R_{\text{available}}^4\propto \frac{\mathcal{V}^{2/3}}{(h^{1,1}_+)^p}\, ,
\end{equation}
for some undetermined positive coefficient $p$. It is easy to see that if we implement the bound on the maximal available dual Coxeter number of eq. \eqref{eq:NvsH11}, one obtains a bound
\begin{equation}
U\gg (g_sM)^2 g_s^{-1}\, \, (h^{1,1}_+)^{p-1}\, . 
\end{equation}
It is apparent that choosing large values of $h^{1,1}_+$ will relax the bound only if the freely available volume scales very weakly with the number of K\"ahler moduli i.e. $p<1$. So we should ask ourselves how large we expect $p$ to be. Instead of asking what is the maximal region around a conifold singularity that is well described by the non-compact conifold solution, it is simpler and arguably less restrictive to ask the following instead. \textit{What is the spherical region of maximal size around a generic point?} We may build a chart $U_p$ around a generic point $p$ so that every point in $\del U_p$ is geodesically equidistant from the center $p$ by a distance $R_p$. Then, what is the largest possible radius $R_p$? Roughly speaking this gives the largest $5$-sphere that can be expanded around a generic point. We do not know how to answer this question for CYs but we expect that generically the size $R_{available}$ available for fitting a warped throat can be bounded by this.

Let us now consider a somewhat different but related question. For full moduli stabilization to occur we expect that each divisor class has a representative that is wrapped by a seven brane or euclidean $D3$ brane instanton. Our conifold region should not be intersected by any of these divisors. In particular none of the triple intersection points should be contained in the conifold region. So in addition one may ask what is the largest available $5$-sphere that does not contain any of the triple intersection points. We expect that both types of largest possible spheres will be bounded for similar reasons.

\begin{figure}
	\centering
	\begin{tabular}{c c}
		\includegraphics[width=6cm,keepaspectratio]{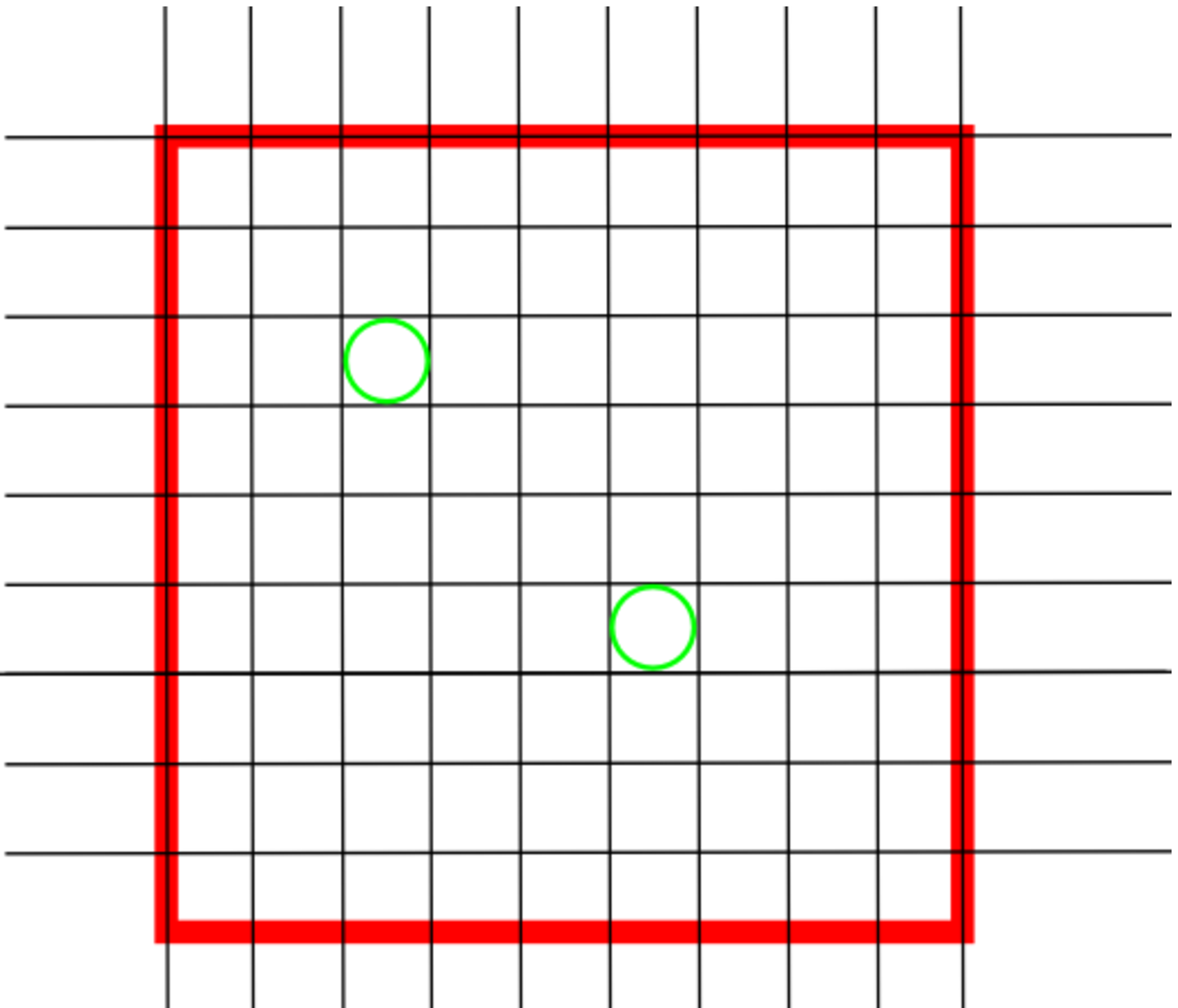} &
		\includegraphics[width=5.1cm,keepaspectratio]{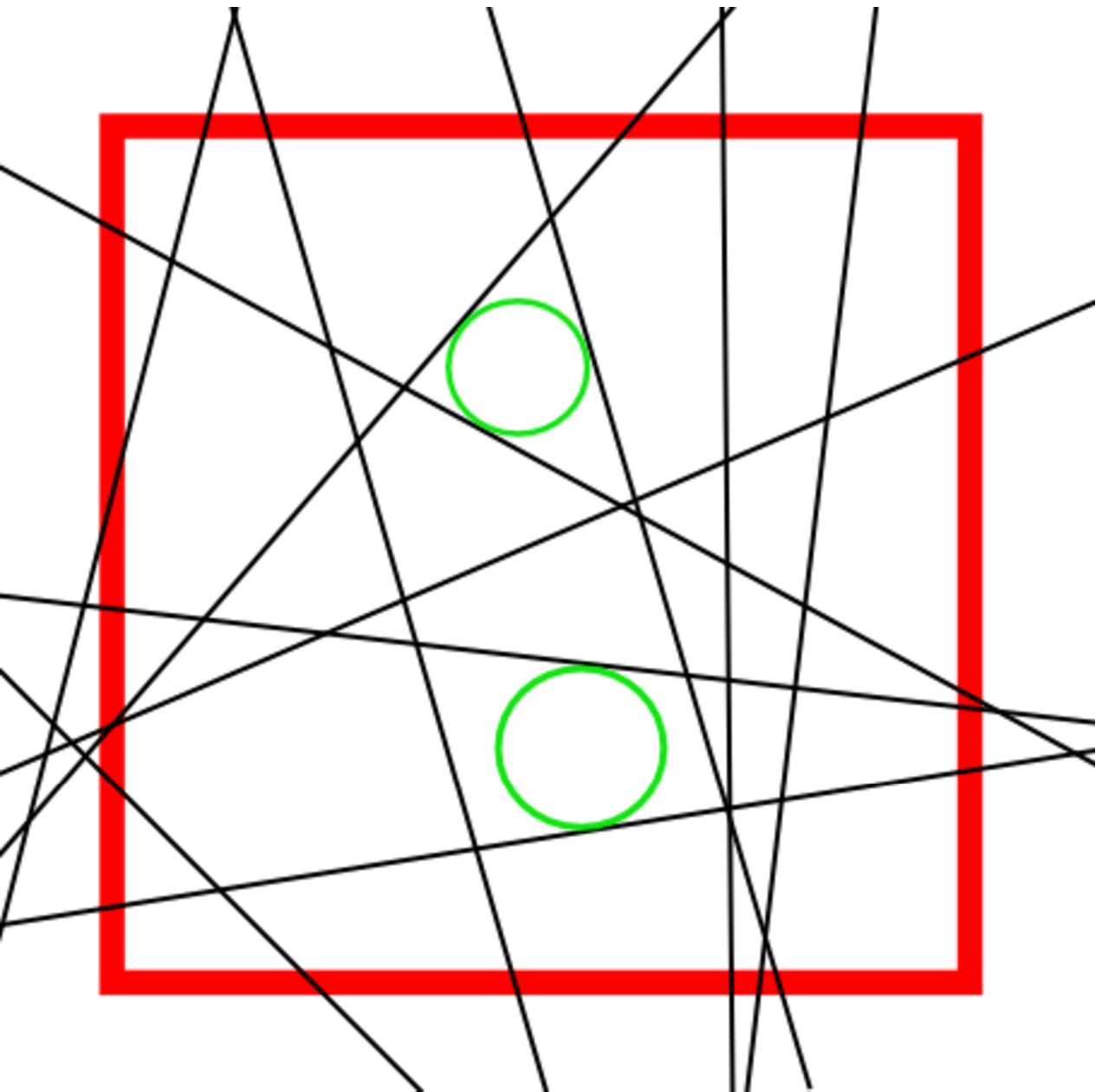}
	\end{tabular}
	\caption{On the l.h. side we plot a square intersected by a grid formed of $n$ lines. The freely available space is of order $1/n^2$. On the r.h. side we show a more generic intersection pattern. We expect the freely available size surrounding a generic point to be of the same order as for the case of a regular grid.}
	\label{fg:sq}
\end{figure}

So let us consider a very simple toy setup where this last question can be easily addressed: Given a six dimensional cube of unit volume, let us randomly intersect it with $n$ real co-dimension $2$ planes. Let us call $R_x$ the radius of the largest $5$-sphere centered at a point $x$ that can be drawn without intersecting any of the intersecting planes (see Figure \ref{fg:sq}). 
Since a generic triplet of intersecting planes intersects at a point we expect $\mathcal{O}(n^3)$ triple intersection points in the cube, distributed randomly. The largest sphere that can be fit must in particular not contain any of these points in its interior so one finds $R_x^6<\mathcal{O}(n^{-3})$
for a generic point $x$. For a cube of overall volume $\mathcal{V}$, this is replaced by
\begin{equation}
R_x^6<\mathcal{O}(n^{-3})\cdot \mathcal{V}\, .
\end{equation}
Obviously cubes intersected by co-dimension $2$ planes are not a good approximation of CYs with wrapped divisors but it may give us some intuition how intersecting co-dimension $2$ branes limit the available un-intersected volume.

In ref. \cite{Demirtas:2018akl} another likewise related question was posed for actual CY three-folds and numerical answers were given using the Kreuzer-Skarke database: \textit{How large does the overall volume of the compactification have to be in order to ensure that the $\alpha'$ expansion is under marginal control?} Numerically, the answer appears to be
\begin{equation}
\text{Vol}(CY)\gtrsim (h^{1,1})^7\, .
\end{equation}
Of course it is tempting to interpret this result as the statement that $p=\frac{2}{3}\cdot 7$. However, the above only means that if all the non-trivial curves are required to be bigger than $\alpha'$ the overall volume must be large. In principle the CY might nevertheless contain large \textit{empty} regions. It would be very interesting to explore how precisely the freely available volume within a CY scales with the number of K\"ahler moduli in order to place the above considerations on a firm footing. 

\begin{figure}
    \centering
    \includegraphics[width=6cm,keepaspectratio]{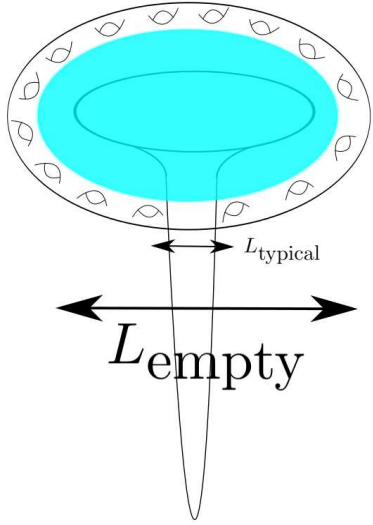}
    \caption{We depict a cartoon of a CY manifold that would evade our conclusions. There exists a large topologically trivial area in the interior where a large KS throat can fit. All of the topological structure is densely aligned around it.}
    \label{fg:Cy4}
\end{figure}

Based on these simplified preliminary observations we find it reasonable to expect that generically it will be hard to engineer $p<1$, while actually proving this to be impossible is beyond the scope of this note. In this case we expect that \textit{generic} CYs do not admit sufficiently small de Sitter uplifts even if $W_0$ and the IR warp factor $a_0^2$ can be tuned at will. Again, we want to stress that non-generic CYs (or CYs at non-generic points in moduli space) might evade this argument. For a cartoon of how such a CY could look like, see Figure \ref{fg:Cy4}. It would be very interesting to see if such non-generic CYs can be engineered for dS uplifts.

As promised, let us consider the potential loophole of exponentially small/large coefficient $A$. It has been explained in ref. \cite{Denef:2005mm} that this is indeed possible due to induced $D_{-1}$-brane charge or induced $D3$ brane charge for the case of euclidean $D3$ brane instantons respectively gaugino condensation on seven branes,
\begin{equation}
|A|\sim e^{\frac{2\pi}{g_sN}\frac{\chi(\Sigma)}{24}}\, .
\end{equation}
This would change our discussion only if one can find solutions where $\log(|A/W_0|)$ is dominated by $\log(|A|)$, so that
\begin{equation}\label{eq:A-dominance}
\text{Re}(T)\sim \frac{N}{2\pi}\log(|A|)\sim \frac{\chi(\Sigma)}{24g_s}
\end{equation}
To us this does not seem to be a controlled regime. This can for instance be seen via inspecting the euclidean $D3$ brane action. To low orders in the $\alpha'$-expansion it takes the form
\begin{equation}
S_{ED3}=\frac{2\pi}{g_s}\left(\underbrace{g_s\text{Re}\,T}_{\text{tree level}}\quad +\quad (-1)\times \underbrace{\frac{\chi(\Sigma)}{24}}_{\mathcal{O}(\alpha'^2)\text{ correction}}\quad +\quad \mathcal{O}(\alpha'^3)      \right)\, .
\end{equation}
In the regime of \eqref{eq:A-dominance} we see that the tree-level contribution is of the same order as the $(\alpha')^2$ correction, indicating loss of control over the  $\alpha'$ expansion.\footnote{This is qualitatively very different to e.g. the KKLT expansion of the superpotential where also the first two terms in the expansion are of the same order. This is achieved via a fine tuning of the first term so one does not expect a breakdown of the non-perturbative expansion scheme.}

As a final and perhaps most interesting loophole, let us note that the KS throat also has a dual description in terms of a confining gauge theory \cite{Klebanov:2000nc,Klebanov:2000hb}. The parameter $g_sM$ sets the 't Hooft coupling of the last steps of the cascade of Seiberg dualities that the gauge theory is undergoing. So the regime $g_sM\ll 1$ is controlled by the gauge theory side of the correspondence. If the SUSY breaking anti-brane state also exists in this regime (and remains meta-stable), the bound on the smallness of the uplift becomes considerably weaker. Even if this holds it is not obvious whether the uplift could then be made sufficiently small. We can interpret the r.h. side of the bound in eq. \eqref{eq:bound_on_N} as the ratio between the size of the IR end of the throat $R_{\IR}^4\sim (g_sM)^2$ and the characteristic ``size" of the anti-brane $R^4_{\overline{D3}}\sim g_s$. Whenever the latter exceeds the former we would expect the brane not to be able to sink down all the way to the bottom of the throat, thus again preventing the uplift potential from becoming small. Whether or not this (after all geometric) intuition carries over to the small $g_sM$ regime remains to be seen. If these considerations are valid, it seems possible that the r.h. side of the bound of eq.~\eqref{eq:bound_on_N} could take $\mathcal{O}(1)$ values. In this case, moderately large rank gauge groups $N
\sim 10$ may be enough to marginally fulfill all constraints, though never parametrically.\footnote{We thank M. Reece for discussions on this point.}

\section{Conclusions and discussion}\label{sec:conclusions}

In this note we have commented on two independent questions. First, we have argued that ten dimensional consistency requirements in the form of tadpole cancellation constraints can be readily satisfied by uplifted KKLT type solutions, under the assumption that the supersymmetric starting point is consistent, and that sufficiently strongly warped throats exist (see section \ref{sec:tadpole_cancellation}). The relevant contributions to the tadpole equations are argued to arise from the stress caused by the restoring force that the stabilization sector exerts on the volume modulus once (say) a warped uplift pulls it towards larger values. We have explained why we find it unlikely that such effects are captured by any local $10d$ action. In particular, they have not been taken into account in the discussions of refs. \cite{Moritz:2017xto,Gautason:2018gln} where it was assumed that an appropriate local action can exist. We would like to emphasize that this is not in conflict with the assertion that \textit{supersymmetric} KKLT vacua can be described in $10d$ using a local action for which there is considerable evidence \cite{Baumann:2006th,Baumann:2010sx,Dymarsky:2010mf,Hebecker:2018vxz,Kallosh:2019oxv} (see also our section \ref{sec:Gauginos} for very basic evidence in this direction).

The second part of this note \ref{sec:warped_uplifts} is largely independent of the first. We have commented on the question if sufficiently long warped throats can exist that could support an uplift to dS (but not overuplift into a run-away solution). We have found that there is a tension between realizing this and also ensuring that the supersymmetric starting point lies within the regime of a controlled $10d$ flux compactification. For the case of a single K\"ahler modulus this tension is resolved only if extremely large gauge group ranks can be realized while there is independent evidence that this cannot be achieved \cite{Louis:2012nb}. We speculate that for the case of many K\"ahler moduli the tension generically becomes stronger while it is possible that CYs with very non-generic geometrical properties can resolve the tension. It would be very interesting to see if such non-generic CYs can be engineered, and if other ideas for uplifting
\cite{Roupec:2018mbn,Flauger:2008ad,Danielsson:2011au,Silverstein:2001xn,Maloney:2002rr,Dodelson:2013iba,Saltman:2004jh,Polchinski:2009ch,Dong:2010pm,Dong:2011uf,Silverstein:2007ac,Burgess:2003ic,Balasubramanian:2004uy,Conlon:2005ki,Westphal:2006tn,Cicoli:2012fh,Louis:2012nb,Cicoli:2013cha,Rummel:2014raa,Cicoli:2015ylx,Braun:2015pza,Retolaza:2015nvh,Gallego:2017dvd}, or other mechanisms of moduli stabilization such as the Large Volume Scenario \cite{Balasubramanian:2005zx} (LVS) can be constrained in a similar manner. We leave these questions for future research.

\bigskip

\centerline{\bf \large Acknowledgments}

We would like to thank Arthur Hebecker, Liam McAllister, Matthew Reece, Ander Retolaza and Thomas Van Riet for useful discussions. This work was supported by the ERC Consolidator Grant STRINGFLATION under the HORIZON 2020 grant agreement no.
647995.

\appendix
\section{The volume modulus in warped compactifications}\label{app:volume_modulus}
In this appendix we recall what is the physical significance of the volume modulus in a warped compactification of the GKP type as explained in \cite{Giddings:2001yu,deAlwis:2003sn,Giddings:2005ff,Shiu:2008ry}. The Einstein frame $10d$ metric of a one-parameter family of solutions to the equations of motion take the form
\begin{align}
ds^2=&t^{-1}e^{2A}dx^2+\alpha'e^{-2A}ds^2_{CY}\, ,\quad e^{-4A}\equiv e^{-4A_0}+(t-t_0)\, ,
\end{align}
where $t_0\equiv \int_{M_6}\sqrt{g_{CY}}e^{-4A_0}$, and $e^{2A_0}$ is some reference solution for the warp factor. The metric $ds^2_{CY}$ is a CY metric normalized to unit volume (or more generally a solution coming from F-theory). $t$ is the real modulus of the solution, and at sufficiently large values the metric approaches
\begin{equation}
ds^2\longrightarrow t^{-3/2}dx^2+\alpha't^{1/2}ds^2_{CY}\, ,
\end{equation}
so in this regime we may identify $t^{3/2}$ with the compactification volume in units of $\alpha'$.

\begin{figure}
	\centering
	\includegraphics[width=12cm,keepaspectratio,angle=270]{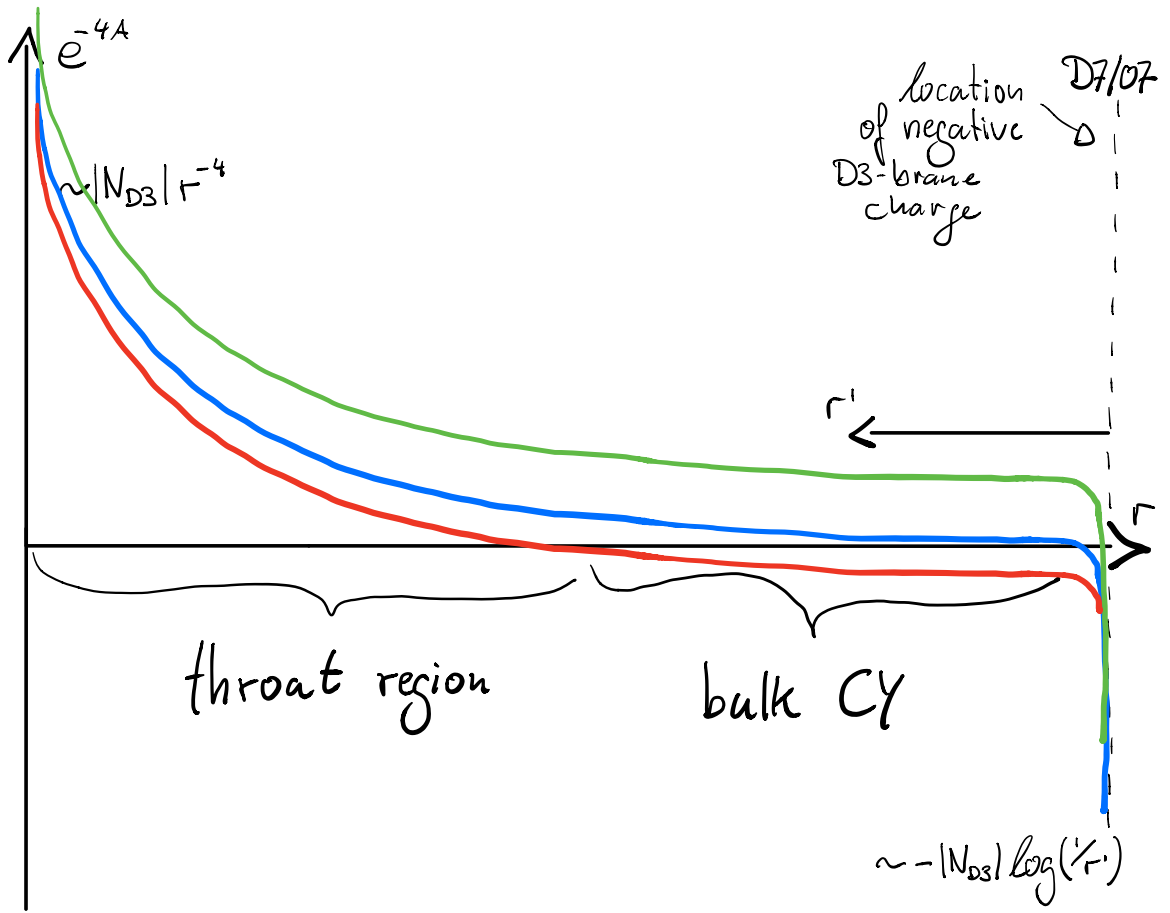}
	\caption{We plot the schematic form of the inverse warp factor in a flux compactification for different values of the volume modulus. It diverges as $|N_{D3}|r^{-4}$ near the position of localized (or approximately localized) positive $D3$ brane charge, and as $-|N_{D3}|\log(1/r')$ near the locus of negative induced $D3$ brane charge on $D7/O7$ stacks. $r$ is the transversal distance to the positive charge while $r'$ is the transversal distance to the negative charge. The blue curve corresponds to taking $t\gtrsim N_{D3}$ where the vanishing locus of $e^{-4A}$ is marginally aligned with the $D7/O7$ loci. The green curve corresponds to the case $t\gg N_{D3}$ where the inverse warp factor vanishes only very close to the seven-brane stacks. The $10d$ geometric picture is under parametric control. In the opposite regime $t< N_{D3}$, shown in red, the singular locus reaches far into the bulk.}
	\label{fg:warping}
\end{figure}

In general, for a GKP type solution, the warp factor is determined by solving a $6d$ Laplace equation
\begin{equation}
\nabla^2_{CY}e^{-4A}\propto \rho_{D3}\, ,
\end{equation}
where $\rho_{D3}$ is the $D3$ brane charge density as measured using the CY metric $g_{mn}^{CY}$ \cite{Giddings:2005ff}. Hence, e.g. near a point like (or smeared along a real co-dimension two locus) source of $N_{D3}$ units of $D3$ brane charge the solution takes the form
\begin{equation}
e^{-4A}\sim \begin{cases}
N_{D3}r^{-4} +const. & \text{near a point-like source} \\
N_{D3}\log(1/r) + const. & \text{near a co-dimension two source}
\end{cases}\, ,
\end{equation}
where $r$ measures the transversal distance to the source (using the dimensionless CY metric). We are interested in cases where the negative $D3$ brane charge is effectively smeared over $D7/O7$ stacks while the positive charge is stored in the fluxes of a KS throat (in such a way that the overall $D3$ brane charge vanishes). W.l.o.g. we may assume that the particular solution $e^{-4A}=e^{-4A_0}$ corresponds to the case where the overall volume is sufficiently small (and not much smaller) that backreaction from $D3$ brane charge cannot be neglected anywhere, but the vanishing locus of the inverse warp factor is still marginally aligned with the loci that carry the negative $D3$ brane charge (see figure \ref{fg:warping}). 
If we now use the one-parameter freedom of the GKP solutions to set
\begin{equation}
e^{-4A}=e^{-4A_0}+(t-t_0)\, ,
\end{equation}
we see that the vanishing locus of $e^{-4A}$ merges into the location of negative $D3$ brane charge as we take $t-t_0\gg N_{D3}$, while for $t-t_0<0$ the vanishing locus quickly moves into the bulk and the $10d$ solution becomes pathological. 

Moreover by inspecting e.g. the solution near a stack of $D3$ branes it is easy to see that $t_0\gtrsim N_{D3}$. Therefore, in order for a controlled $10d$ solution to exist where all pathological behavior is concentrated close to the seven-brane stacks, we need $t\gg N_{D3}$. In this case the physical volume of the CY is also much bigger than $N_{D3}^{3/2}$ and well approximated by the value of $\alpha'^3t^{3/2}$. This is also the regime where we may think of warped throats as isolated regions of strong warping embedded into an essentially trivially warped bulk CY.

\newpage
\bigskip

\bibliographystyle{JHEP}
\bibliography{RefsUplift.bib}

\end{document}